\documentclass[letterpaper]{JHEP3}

\usepackage{epsfig,multicol,amsmath}

\newcommand{\roughly}[1]{\mathrel{\raise.3ex\hbox{$#1$\kern-0.85em
\lower1ex\hbox{$\sim$}}}}

\newcommand{\lsim}{\roughly<}

\def\exd{{\hbox{d}}}

\def\endignore{}
\def\ignore #1\endignore{} 

\def\ba{\begin{eqnarray}}
\def\ea{\end{eqnarray}}
\def\be{\begin{equation}}
\def\ee{\end{equation}}

\def\ssA{{\scriptscriptstyle A}}

\def\AcB{{\scriptscriptstyle A,B}}
\def\ssB{{\scriptscriptstyle B}}

\def\NR{{\scriptscriptstyle NR}}

\def\bfx{{\bf x}}
\def\bfz{{\bf z}}

\def\F{\mathcal{F}}
\def\G{\mathcal{G}}

\def\L{\mathcal{L}}

\def\O{\mathcal{O}}
\def\R{\mathcal{R}}

\def\nn{\nonumber}

\def\({\left(}
\def\){\right)}

\def\pref#1{(\ref{#1})}

\title{Cosmic Black-Hole Hair Growth and Quasar OJ287}

\author{M.W.~Horbatsch${}^1$ and C.P.~Burgess${}^{1,2}$\\

$^1$ Dept. of Physics \& Astronomy, McMaster University \\
 \qquad 1280 Main St. W, Hamilton, Ontario, Canada, L8S 4L8.\\
\\
$^2$ Perimeter Institute for Theoretical Physics \\
 \qquad 31 Caroline St. N, Waterloo, Ontario, Canada  N2L 2Y5.\\
}

\abstract{An old result ({\tt astro-ph/9905303}) by Jacobson implies that a black hole with Schwarzschild radius $r_s$ acquires scalar hair, $Q \propto r_s^2 \mu$, when the (canonically normalized) scalar field in question is slowly time-dependent far from the black hole, $\partial_t \phi \simeq \mu M_p$ with $\mu r_s \ll 1$ time-independent. Such a time dependence could arise in scalar-tensor theories either from cosmological evolution, or due to the slow motion of the black hole within an asymptotic spatial gradient in the scalar field. Most remarkably, the amount of scalar hair so induced is independent of the strength with which the scalar couples to matter. We argue that Jacobson's Miracle Hair-Growth Formula${}^\copyright$ implies, in particular, that an orbiting pair of black holes can radiate {\em dipole} radiation, provided only that the two black holes have different masses. Quasar OJ 287, situated at redshift $z \simeq 0.306$, has been argued to be a double black-hole binary system of this type, whose orbital decay recently has been indirectly measured and found to agree with the predictions of General Relativity to within $6\%$. We argue that the absence of observable scalar dipole radiation in this system yields the remarkable bound $|\,\mu| < (16 \, \hbox{days})^{-1}$ on the instantaneous time derivative at this redshift (as opposed to constraining an average field difference, $\Delta \phi$, over cosmological times), provided only that the scalar is light enough to be radiated --- {\em i.e.} $m \lsim 10^{-23}$ eV --- independent of how the scalar couples to matter. This can also be interpreted as constraining (in a more model-dependent way) the binary's motion relative to any spatial variation of the scalar field within its immediate vicinity within its host galaxy.
}

\begin{document}

\section{Introduction}

Ten years ago Ted Jacobson discovered Jacobson's Miracle Hair Growth Formula \cite{jacobson}, which shows how black holes can acquire scalar hair if they are embedded into regions in which the relevant scalar fields evolve slowly with time. We argue in this paper that this observation can significantly change the behaviour of black-hole binaries, and provide very robust new observational constraints on the existence of time-dependent scalar fields in the cosmological past. Unlike other recent efforts to constrain scalar-tensor theories using black hole physics \cite{axiverse}, our arguments do not rely on black-hole super-radiance.

Why consider light scalars? Light scalar fields provide one of the few known ways to consistently modify gravity over large distances, and one of the main ways in which candidate quantum theories of gravity can differ from one another is the spectrum of bosons they predict at very low energies. Such fields, if they exist and are sufficiently light, can mediate long-range forces that can observably compete with gravity. Because of this many of their properties have been carefully studied over the years \cite{damourrev, sctensrev, BD, otherlightsc, chameleon, galileon}.

Furthermore, scalars commonly arise in fundamental theories, although it is more unusual for them to be light enough to mediate forces over macroscopic distances. Scalars are rarely this light because quantum corrections famously tend to give large contributions to scalar masses, even if they would have been massless at the purely classical level. But in some circumstances symmetries can protect against masses, such as if the scalar is a pseudo-Goldstone boson \cite{pGB} for a spontaneously broken approximate symmetry. This is what protects the mass of an axion \cite{axionorig,axionrev}, or other axion-like fields \cite{moreaxion}. Alternatively, light scalars can arise as part of the low-energy limit of an extra-dimensional model, with mass protection coming from extra-dimensional symmetries like higher-dimensional general covariance \cite{SLEDCosmo, ubernat}. In particular, extremely light scalars with masses of order the present-day Hubble scale naturally arise in this way in extra-dimensional proposals to address the cosmological constant problem \cite{SLED, SLEDrev}.

Once a light scalar is present it is very unlikely to be time-independent, since a small potential can easily drive a light scalar to roll. So it is natural to examine how the time evolution of ambient scalar fields can affect the properties of black holes and other astronomical objects. It is in this context that Jacobson's observation of scalar hair growth by black holes is so interesting.

We find in this paper that the scalar hair Jacobson predicts for black holes can have potentially measurable implications, by modifying the post-Newtonian response of orbiting black holes. In particular, we find it provides an unexpected source of scalar dipole radiation for an orbiting black hole, which would have been unavailable if the black hole had no hair (as is usually supposed).

Specifically, for canonically normalized scalar fields that approach $\phi \to \phi_\infty + \mu t$ asymptotically far from a pair of orbiting near-Newtonian black holes (with masses $M_\ssA$ and $M_\ssB$), we find the fraction of power radiated into scalar fields to be proportional to $\Delta a^2$, where $\Delta a = 4 G \mu (M_\ssA - M_\ssB)$. We apply this expression to the quasar OJ287, which has been interpreted as a double black hole binary, leading to a bound $|\, \mu| < (16 \, \hbox{days})^{-1}$ on the instantaneous time-variation of {\em any} sufficiently light (see below for precisely how light) scalar fields at redshift $z \simeq 0.3$.

Others \cite{OtherJMHGF} have also sought useful ways to apply Jacobson's Miracle Hair Growth Formula, through its potential implications for black-hole accretion \cite{BHaccretion}. Rather than using (as we do) the scalar hair to argue for dipole scalar radiation, these authors argue that in quintessence models primordial black holes can accumulate significant amounts of the quintessence field, potentially leading to observable effects in gravitational radiation.

This paper is organized as follows. In section \ref{sec:sctensbhb}, scalar-tensor gravity is introduced, and the no-hair theorem is explicitly illustrated for the static spherically-symmetric solutions of the vacuum field equations. In section \ref{sec:massschair}, the mass $M$ and scalar charge $Q$ of an astrophysical body are defined, in terms of both the behaviour of the fields asymptotically far from the body, and the functional derivatives of the point-particle action of the effective field theory in which the internal structure of the body has been integrated out. In section \ref{sec:bh}, Jacobson's time-dependent black hole solution with scalar hair (the Miracle Hair Growth Formula) is described. An analysis of the appropriate point-particle action is carried out, and it is found that this type of black hole couples to the scalar field differently than a star with the same scalar charge, implying that the standard formulas in the scalar-tensor literature for the Post-Keplerian orbital corrections need not apply to a binary system of two such black holes. Section \ref{sec:bhbinfr} then argues that standard formulae {\em do} apply for the leading-order rate of scalar dipole radiation emitted by a such a binary system. In section \ref{sec:oj287}, the result of this calculation is applied to the quasar OJ287. Finally, section \ref{sec:summary} summarizes our results and prospects for future work.

\section{Black holes: hair and dipole radiation}
\label{sec:bhhairdip}

This section contains the main formal results: describing black holes in scalar-tensor theories, with both static and slowly varying asymptotic scalar configurations. It also briefly summarizes the standard formulae for the dipole and quadrupole radiation rate from binary pairs in the near-Newtonian limit.

\subsection{Scalar-Tensor gravity and black-hole balding}
\label{sec:sctensbhb}

The simplest variants of scalar-tensor gravity involve a single scalar field, $\psi$, and the space-time metric, $\hat g_{\mu\nu}$, with action: $S = S_g + S_{\rm mat}$. Here $S_{\rm mat}$ describes how the fields $\psi$ and $g_{\mu\nu}$ couple to other matter kinds of matter and $S_g$ describes the mutual interactions of $\psi$ and the metric. Writing $S_g$ in a derivative expansion and keeping only up to two derivatives, the most general expression is
\begin{equation}
\label{actionv0.9}
 S_{g} = - \int \exd^{4}x
 \sqrt{-\hat g} \; \left\{ U(\psi) + \hat g^{\mu \nu}
 \left[ \F(\psi) \, \hat{\mathcal{R}}_{\mu \nu}
 + \G(\psi) \, \partial_{\mu} \psi \,
 \partial_{\nu} \psi \right] + \cdots \right\}  \,,
\end{equation}
where the form taken by the functions $U(\psi)$, $\F(\psi)$ and $\G(\psi)$ characterize a particular scalar-tensor theory. Here $\hat{\mathcal{R}}_{\mu\nu}$ denotes the Ricci tensor\footnote{Conventions: we use metric signature $(-+++)$ and Weinberg's curvature conventions \cite{Wbg} (differing from MTW \cite{MTW} only by an overall sign for the Riemann tensor). Unless specified otherwise, units are chosen throughout with $\hbar = c = k_\ssB = 1$.} constructed from the metric $\hat g_{\mu\nu}$. Of particular interest are scalar-tensor theories which satisfy the weak equivalence principle. This can be ensured by requiring that matter couplings, $S_{\rm mat}$, depend only on $\hat g_{\mu\nu}$ and not separately on $\psi$.

In practice the functions $\F$ and $\G$ need not be specified in detail because they can be dramatically simplified by performing appropriate field redefinitions of the form $\hat g_{\mu\nu} = A^{2}(\psi) \, g_{\mu\nu}$ and $\psi = \psi(\varphi)$, with the result
\begin{equation}
\label{action}
 S_{g} = -\frac{1}{16 \pi G}\int \exd^{4}x \sqrt{-g}
 \; \Bigl[ g^{\mu \nu} \left( \mathcal{R}_{\mu \nu}
 + 2 \, \partial_{\mu} \varphi \,
 \partial_{\nu}\varphi \right) + 16 \pi G \, V(\varphi)
 + \cdots \Bigr] \,,
\end{equation}
where $G$ is Newton's constant. This leaves $V(\varphi)$ as the only important function governing the long-distance properties of the scalar-gravity couplings. For those theories for which $S_{\rm mat}$ depends only on $\hat g_{\mu\nu} = A^2(\varphi) g_{\mu\nu}$ and not separately on $\varphi$, all of the direct matter couplings of the field $\varphi$ are described by the function $A(\varphi) := A[\psi(\varphi)]$. For such theories the two functions $A(\varphi)$ and $V(\varphi)$ completely define the low-energy predictions. Brans-Dicke theories \cite{BD} comprise the widely studied special case where $V(\varphi) = 0$ and $A(\varphi) = e^{a \, \varphi}$, for constant $a$.

In what follows we require no assumptions about the scalar-matter couplings, but our interest is in situations where the scalar potential, $V(\varphi) = \left. A^4(\psi) \, U(\psi) \right|_{\psi(\varphi)}$, is negligible (making $\varphi$ a minimally coupled massless scalar). The vacuum field equations obtained from (\ref{action}) when $V \simeq 0$ are very simple
\begin{eqnarray}
\label{fieldeq_tens}
 \mathcal{R}_{\mu \nu}+ 2 \, \partial_{\mu} \varphi \, \partial_{\nu} \varphi  &=&   0 \,, \\
 \label{fieldeq_sc}
 \Box \, \varphi = \frac{1}{\sqrt{-g}} \, \partial_\mu
 \left( \sqrt{-g} \; g^{\mu\nu} \partial_\nu \varphi
 \right) &=&  0 \,.
\end{eqnarray}

Time-independent, spherically symmetric solutions to these equations are known \cite{spheresoln, damourrev, BMQ}, and given by
\be
 \exd s^2 = - f(r) \, \exd t^2 + \frac{\exd r^2}{f(r)}
 + g(r) \, \Bigl( \exd \theta^2 + \sin^2 \theta \,
 \exd \phi^2 \Bigr)
 \quad \hbox{and} \quad
 \varphi = \varphi(r) \,,
\ee
where
\ba \label{sphsymsolns}
 f(r) &=& \left( 1 - \frac{\ell}{r} \right)^\alpha \nn\\
 g(r) &=& r^2 \left( 1 - \frac{\ell}{r} \right)^{1-\alpha} \\
 \hbox{and} \quad
 \varphi(r) &=& \varphi_\infty +
 \frac{q}{2} \ln \left( 1 -
 \frac{\ell}{r} \right) \,,\nn
\ea
and the constants $\alpha$ and $q$ must satisfy $\alpha^2 + q^2 = 1$.

The Schwarzschild black hole corresponds to the choice $\alpha = 1$ and $q = 0$, in which case $\varphi = \varphi_\infty$ is constant, and the constant $\ell$ is revealed as the Schwarzschild radius: $\ell = 2 GM$. In this context the no-hair theorems \cite{nohair} state that $\alpha = 1$ and $q = 0$ is the only physically allowed black-hole solution. Within the family defined by eqs.~\pref{sphsymsolns}, the problem with the more general solutions is that $\varphi$ always diverges at $r = \ell$ whenever $q \ne 0$. Since the field equation \pref{fieldeq_tens} shows that the curvature scalar, $\R$, also diverges there, $r = \ell$ is an honest-to-God curvature singularity, rather than simply marking the coordinate singularity associated with an event horizon.

\subsection{Masses and scalar hair}
\label{sec:massschair}

We next pause to define scalar hair more precisely, for use in later sections. One way to define the (ADM) mass \cite{ADM, BMQ}, $M$, and the `scalar charge' \cite{damourrev, Stell1, StellUS}, $Q$, is through the large-$r$ asymptotic form of the metric and the scalar field\footnote{Notice our conventions here for $Q$ follow those of ref.~\cite{StellUS}, and so differ by a sign from those of refs.~\cite{damourrev, Stell1}, who define $\varphi = \varphi_\infty + G\omega/r + \cdots$.}
\be
 g_{tt} = -1 + \frac{2 G M}{r} + \cdots
 \quad \hbox{and} \quad
 \varphi = \varphi_\infty - \frac{GQ}{r} + \cdots
 \,.
\ee
In terms of these definitions we see that the integration constants $\ell$ and $q$ can be traded for $M$ and $Q$, with
\be
 2GM = \alpha \, \ell \quad \hbox{and} \quad
 GQ = \frac{q \ell}{2} \,.
\ee

For later purposes it is worth elaborating a bit on the extent to which this definition of $Q$ provides a useful definition of `charge'. The main observation is that the scalar equation of motion, eq.~\pref{fieldeq_sc}, can be read as a conservation law: $\partial_\mu J^\mu = 0$, for $J_\mu := \sqrt{-g} \; \partial_\mu \varphi$. Although the quantity $Q$ is {\em not} the charge --- {\em i.e.} the spatial integral of $J^t$ --- for this conserved current, it is `conserved' in the sense that eq.~\pref{fieldeq_sc} implies that the combination $\sqrt{-g} \; g^{rr} \varphi'$ is independent of $r$ (with $\varphi' := \partial_r \varphi$), for any static spherically symmetric configuration. $Q$ as defined above is equivalent to this $r$-independent quantity because
\be
 \sqrt{-g} \;  g^{rr} \varphi' = f \, g \, \varphi' \,
  \sin\theta = \left( 1 - \frac{\ell}{r} \right) r^2 \varphi' \sin \theta = \frac{q \ell}{2} \; \sin \theta = GQ \, \sin \theta \,.
\ee

Our interest in the rest of this paper is in applications computing the radiation rate and post-Newtonian corrections to orbits, for which it is only the centre-of-mass motion of the black hole that is physically relevant. Furthermore, all objects are located far enough from one another to allow their weak mutual interaction to be described by the far-field regime. In this case it is more useful to define mass and charge in terms of the effective point-particle action of the black hole which governs the evolution of its centre of mass \cite{PPAction, DamEFGW, NRCMEFT, NRCMEFTrev},
\be \label{PPactiondef}
 S_p := \int_\Gamma \exd \tau \,
 \L_p[\varphi, g_{\mu\nu}, \partial_\alpha \varphi, R_{\alpha\beta\gamma\delta}, \dot z, \cdots]\,,
\ee
where $\L_p$ is a scalar whose dependence on the background metric and $\varphi$, and their derivatives depends on the physical properties of the object of interest ({\em e.g.} planet, star, neutron star, black hole). This dependence is obtained (see below) by matching the physical predictions of this effective theory to the predictions of a more microscopic description of the object in question \cite{damourrev, NRCMEFT}. In eq.~\pref{PPactiondef} $\L_p$ is evaluated along the point-particle world-line, $z^\mu(\tau)$, denoted as $\Gamma$, with $\tau$ representing proper time (arc-length) measured along $\Gamma$. ($\L_p$ could also depend on other variables for objects with internal degrees of freedom, like spin.) Over-dots denote the derivative with respect to $\tau$, as in $\dot z^\mu := \exd z^\mu/\exd \tau$.

This action has two important uses. First, it can be used as a source in the field equations for $\varphi$ and $g_{\mu\nu}$, and expressions for $\L_p$ can be chosen to make the resulting solutions agree with the far-field solutions computed from an explicit description of the relevant object. (For example, for a star these solutions can be compared to the far-field solutions obtained by matching to interior solutions to the equations of hydrostatic equilibrium inside the star.)

Second, its variation with respect to $z^\mu(\tau)$ can be used to determine the trajectory taken by the object moving within its local geometric environment, including any applied fields $\varphi$. For these latter applications it is useful to organize $\L_p$ into an expansion in derivatives of the background fields,
\be \label{higherderivsv0}
 S_p = \int_\Gamma \exd \tau \Bigl[ m(\varphi) + h(\varphi) \,
 \dot z^\mu \dot z^\nu \, \partial_\mu \varphi
 \, \partial_\nu \varphi + k(\varphi) \, \partial_\mu \varphi \, \partial^\mu \varphi  + \cdots \Bigr] \,,
\ee
since it is often only the first term, $\L_p \simeq m(\varphi)$ that is relevant for practical applications  \cite{DamEFGW}. The coefficients like $m$, $h$ and $k$ take values whose order of magnitude can be determined on dimensional grounds in terms of the mass, $M$, size, $L$, and time-scale $T$ of the source in question; typically with $m \sim \O(M)$, and $k \sim \O(M L^2)$ and $h \sim \O(M T^2)$ and so on. (Here the characteristic time-scale is $T \sim L/v$, where $v$ denotes the typical speed of objects within the source; $v \simeq c = 1$ for relativistic sources like neutron stars.) For practical applications to post-Newtonian orbits it is usually the non-relativistic limit of the point-particle action evaluated in the field of the other body that is relevant, which we obtain by expanding in powers of $v$. (Although the systematics of this expansion are interesting in their own right \cite{DamEFGW,NRCMEFT,NRCMEFTrev} we do not follow these complications here.)

Our immediate interest is in constraining the function, $\L_p(\varphi)$, that describes the black-hole source that gives rise to the explicit scalar-tensor solutions given by eqs.~\pref{sphsymsolns}, above. This may be accomplished by solving the field equations in the presence of the point-particle action, $\delta (S_{g} + S_{p}) = 0$, for instance
\be \label{scaleq_pp}
 \partial_\mu \Bigl( \sqrt{-g} \; g^{\mu\nu} \partial_\nu
 \varphi \Bigr) + 4\pi G \left( \frac{\delta \L_p}{\delta \varphi} \right) \;  \delta^3[x - z(s)] = 0 \,,
\ee
and comparing the far-field result to the solutions of eqs.~\pref{sphsymsolns}. We do not use the derivative expansion, eq.~\pref{higherderivsv0}, when doing so because this can break down when evaluating $S_p$ at self-field configurations sourced by the object itself, due to the divergence of these self-fields at the position of the source.\footnote{These divergences arise already at the classical level, and can be handled systematically by renormalizing the particle action \cite{PPAction, DamEFGW, NRCMEFT, NRCMEFTrev}, as is also relevant in other contexts \cite{BraneRenorm}.}

Taking a static source, and integrating eq.~\pref{scaleq_pp} at a fixed time over a spherical region centered on the source position implies in particular
\be \label{QvsLp}
 -G \, \left. \frac{\delta \L_p}{\delta \varphi}
 \right|_p = \frac{1}{4\pi}
 \oint_{r,t} \exd \theta \exd \phi \, \sqrt{-g} \; g^{rr} \varphi' =  f(r) \, g(r) \, \varphi'(r)
 =  GQ \,,
\ee
where the far-left-hand term is evaluated at the source position (which generically diverges and so must be renormalized). The Einstein equations give a similar expression involving the stress energy derived from $\L_p$. For example, for widely separated, weakly interacting objects moving within a field that asymptotes at large $r$ to a static constant configuration, $\varphi_\infty$, and assuming $\L_p \simeq m(\varphi)$, these considerations reproduce the standard results \cite{damourrev}
\be \label{MQvsm}
 M = m(\varphi_\infty)
  \quad \hbox{and} \quad
  Q = -m'(\varphi_\infty) \,.
\ee

Notice that use of the weak, far-field expansion does {\em not} imply any assumptions about the nature of the underlying source, and in particular does not exclude it being a relativistic star or a black hole \cite{damourrev}. However, in the special case that the source {\em is} constructed purely from non-relativistic matter, and when this matter couples to $\varphi$ only through the Jordan-frame metric, $\hat g_{\mu\nu}$, we would additionally know that $m(\varphi) \, \exd \tau = m_0 \, \exd \hat \tau$, where $\hat \tau$ is measured using the metric $\hat g_{\mu\nu}$. In this case
\be
 m_\NR(\varphi)
  \simeq m_0 \, A(\varphi) \,,
\ee
leading to $M_\NR \simeq m_0 \, A(\varphi_\infty)$ and
\be \label{NRQMrel}
 Q_\NR \simeq -m_0 A'(\varphi_\infty) = -M_\NR \, a(\varphi_\infty) \,,
\ee
where $a(\varphi) := A'(\varphi)/A(\varphi)$ \cite{damourrev}.
Eq.~\pref{NRQMrel} agrees with more detailed studies of stellar structure in scalar-tensor gravity \cite{Stell1,StellUS}, which also show how $Q/M$ differs from $a(\varphi_\infty)$ as the underlying stellar equation of state becomes more relativistic.

What is the function, $m(\varphi)$, appropriate when the object in question is a nonsingular, static black hole? From solutions eqs.~\pref{sphsymsolns} we see that static black hole solutions have a constant scalar profile, $\varphi = \varphi_\infty$, on which the asymptotic metric does not depend. Consequently $M = m(\varphi_\infty)$ is independent of $\varphi_\infty$ --- which is consistent with $Q = -m'(\varphi_\infty) = 0$, as found above.

For black holes the conclusion that $m(\varphi)$ must be independent of $\varphi$ should be very general, and should also apply to the coefficients $h(\varphi)$ and $k(\varphi)$ of the expansion, eq.~\pref{higherderivsv0}. It is this general because it is not necessary to couple to matter at all in order to form a black hole, and so the effective action for a black hole should be invariant under the shift symmetry, $\varphi \to \varphi + \hbox{constant}$, of the classical field equations, eqs.~\pref{fieldeq_tens} and \pref{fieldeq_sc}.

\subsection{Jacobson's Miracle Hair-Growth Formula}
\label{sec:bh}

For massless scalars the asymptotic value, $\varphi_\infty$, of the scalar field can be chosen freely, and is not related to the physics of a star or black hole. This is easily understood because far from the black hole the geometry becomes flat, and any constant value $\varphi = \varphi_\infty$ solves the scalar field equation, $\Box \,\varphi = 0$. However a function linear in time, $\varphi = \mu t$, is another solution to $\Box \, \varphi = 0$ for flat spacetime, and so one might ask what the black hole solution might be that asymptotes to this for large $r$.

This question need not be completely academic. For instance, for cosmological applications the scalar potential, $V(\varphi)$, need not precisely vanish and might drive a time evolution for $\varphi$ far from the black hole. For small $V$ this evolution could be adiabatic on the time-scales relevant to black-hole physics, and so be well-approximated by a slow, approximately linear, variation of $\varphi$ at infinity.

Although the exact solution to this problem is not known, a perturbative solution to eqs.~(\ref{fieldeq_tens}) and (\ref{fieldeq_sc}) is known with the property that $\varphi \to \varphi_\infty + \mu t$ as $r \to \infty$ \cite{jacobson}. This solution is given as an expansion about the black-hole solution,
\begin{eqnarray}
 g_{\mu \nu} &=& g_{\mu \nu}^{(0)} + \epsilon^{2} g_{\mu \nu}^{(2)} + \ldots \,,
 \nonumber
 \\
 \label{pertexp}
 \varphi &=& \varphi^{(0)} +
 \epsilon \varphi^{(1)} + \epsilon^{2} \varphi^{(2)} + \ldots \,,
\end{eqnarray}
where $\epsilon$ is a small dimensionless parameter and
\begin{equation}
\label{schwarzschild}
 g_{\mu \nu}^{(0)} \exd x^{\mu} \exd x^{\nu} =
 - \left( 1 - \frac{2GM}{r} \right)
 \, \exd t^{2} + \frac{\exd r^{2}}{1-2GM/r} + r^{2} \exd\Omega_{2}^{2}
\end{equation}
is the Schwarzschild metric. $\varphi^{(0)} = \varphi_\infty$ is a constant (which we can take to vanish with no loss of generality), and
\begin{equation}
\label{jacobson_sol}
 \varphi^{(1)}_{\pm} = \frac{t}{2GM} \pm \log \left|
 1 - \frac{2GM}{r} \right|
\end{equation}
is the leading scalar-field perturbation. Both choices of sign solve eq.~(\ref{fieldeq_sc}), but we will shortly see that only $\varphi_{+}$ is physically realistic for our later applications. The back-reaction of the metric first arises at $\mathcal{O}(\epsilon^{2})$, because equation (\ref{fieldeq_tens}) is quadratic in $\partial \varphi$.

Comparing this solution at large $r$ to $\mu t$ shows that $\mu = \epsilon/2GM$ and so
\begin{equation}
 \varphi_{\pm} = \mu t \pm 2GM\mu \log \left|
 1-\frac{2GM}{r}\right| +
 \mathcal{O}[(GM\mu)^{2}] \,.
\end{equation}
Thus an expansion in $\epsilon \ll 1$ corresponds to choosing the time scale $\mu^{-1}$ over which the asymptotic scalar field $\varphi$ evolves to be very long relative to the light-crossing time scale, $2GM/c^{3}$, of the black hole.

The necessity for the $r$-dependence in $\varphi^{(1)}$ can be seen by asking how the solution behaves near the black-hole event horizon. In ref.~\cite{jacobson} Jacobson shows that the singularity at $r = 2GM$ in the spatial part of $\varphi^{(1)}$ is required to cancel the singularity due to the breakdown of the coordinate $t$ at the horizon. To see this, we re-write it using coordinates that do not break down there, such as
\begin{equation}
 u = t - r_{\star} \,, \qquad
 v = t + r_{\star} \,,
\end{equation}
where
\begin{equation}
 r_{\star} = r + 2GM \log \left| \frac{r}{2GM} - 1 \right|
\end{equation}
is the `tortoise' coordinate. Trading $t$ for $v$, the Schwarzschild metric is
\begin{equation}
 g_{\mu \nu}^{(0)} \exd x^{\mu} \exd x^{\nu}
 = -\left( 1 - \frac{2GM}{r}\right) \exd v^2
 + 2 \, \exd r \,
 \exd v + r^{2} \exd\Omega^{2}_2 \,,
\end{equation}
and the linearized scalar solutions become
\begin{equation}
 \varphi^{(1)}_{+} = \left( \frac{v-r}{2GM} \right)
  - \log \left(
 \frac{r}{2GM} \right) \,,
 \quad
 \varphi^{(1)}_{-} = \left( \frac{u+r}{2GM}
 \right) + \log\left( \frac{r}{2GM} \right) \,.
\end{equation}
Now, for an eternal black hole the future event horizon is at $t \to \infty$ and $r=2GM$, which corresponds to $u \to \infty$ and finite $v$. The past horizon is at $t\to -\infty$ and $r=2GM$, which corresponds to $v\to -\infty$ and finite $u$. Therefore $\varphi^{(1)}_{+}$ is regular at the future horizon and singular at the past horizon, while $\varphi^{(1)}_{-}$ is regular at the past horizon and singular at the future horizon. Thus, it is the perturbation $\varphi_{+}^{(1)}$ that is relevant for a physically realistic black hole formed by gravitational collapse with no past horizon.

\subsubsection*{Quantifying how much hair}

Because $\varphi^{(1)}_+$ has a radial nontrivial profile, it also introduces a nonzero scalar charge. Expanding eq.~(\ref{jacobson_sol}) in powers of $1/r$ yields
\be
 \varphi^{(1)}_{+} = \frac{t}{2GM} - \frac{2GM}{r}
 + \mathcal{O}\left( \frac{1}{r^{2}} \right) \,,
\ee
and so comparing the $1/r$ term with $-GQ/r$ implies
\be
 \frac{Q}{M} = 2\, \epsilon = 4GM \mu \,,
\ee
indicating that asymptotic scalar time-dependence gives black holes nonzero scalar charge.

Another route to the same conclusion is to recognize that so long as we restrict to spherically symmetric field configurations with a time-independent metric and $\partial_t \varphi$ nonzero but $t$-independent, it follows that $\Box \, \varphi = 0$ still implies the $r$-independence of $\sqrt{-g} \; g^{r\mu} \partial_\mu \varphi = f \, g \, \varphi' \, \sin \theta$. Using Jacobson's asymptotic expression for large $r$ then shows its value for all nonzero $r$ must be
\be
  f(r) \, g(r) \, \varphi' = \lim_{r \to \infty} f(r) \, g(r) \, \varphi' = \lim_{r \to \infty} r^2 \varphi' = 2 GM \, \epsilon \,,
\ee
and so $GQ = (2 GM)^2 \mu$, as concluded above. What is important for later post-Newtonian applications is the black hole's dimensionless charge-to-mass ratio,
\begin{equation}
\label{efscmcoupl}
 a := \frac{Q}{M} =  2 \epsilon = 4 GM \mu \,.
\end{equation}
We saw above that for ordinary stars with non-relativistic equations of state in quasi Brans-Dicke theories this quantity would play the role of the effective coupling, $a(\varphi_\infty)$, to the scalar field far from the star. This need no longer be true for relativistic systems like neutron stars or black holes \cite{damourrev,Stell1,StellUS}.

The utility of this last derivation lies in its connection to eq.~\pref{QvsLp}, which shows that the point-particle effective action for such a black hole must satisfy
\be
 -G \, \left. \frac{\delta \L_p}{\delta \varphi}
 \right|_p =  f(r) \, g(r) \, \varphi'(r)
 = (2 GM)^2 \mu \,.
\ee
Notice that this conclusion is not inconsistent with the black-hole shift symmetry, since this symmetry does not preclude a nonzero functional derivative for $S_p$.

\subsubsection*{Orbital corrections}

Where the existence of the shift symmetry {\em does} play an important role is in how the black hole responds to applied scalar and gravitational fields (such as those due to a companion within a binary system). For these applications a derivative expansion is appropriate, and gives
\be \label{higherderivs}
 S_p = \int_\Gamma \exd \tau \Bigl[ m(\varphi) + h(\varphi) \,
 \dot z^\mu \dot z^\nu \, \partial_\mu \varphi
 \, \partial_\nu \varphi + k(\varphi) \, \partial_\mu \varphi \, \partial^\mu \varphi  + \cdots \Bigr] \,,
\ee
where the functions $h(\varphi)$ and $k(\varphi)$ are to be determined by an appropriate matching calculation. A possible term linear in $\dot z^\mu$ --- {\em i.e.} $\int_\Gamma \exd \tau \, f(\varphi) \, \dot z^\mu \, \partial_\mu \varphi$ --- is not written in eq.~\pref{higherderivs} because it is a total derivative and so does not contribute to the local equations of motion, and several other terms involving the background Ricci tensor can be removed by performing a judicious field redefinition \cite{DamEFGW, GREFT}. The ellipses include terms with either more powers of $\partial_\mu \varphi$, derivatives of the metric, or higher derivatives of $\varphi$.

As before, for applications to black holes the functions $m(\varphi)$, $h(\varphi)$ and $k(\varphi)$ are independent of $\varphi$, since black hole solutions do not require couplings between $\varphi$ and matter and so do not break the classical symmetry under constant shifts: $\varphi \to \varphi + \omega$. When calculating the response of the black hole to fluctuations, $\delta \varphi = \varphi - \varphi_\infty - \mu t$, about the asymptotic scalar, the above action shows the leading terms linear in $\delta \varphi$ are
\ba \label{NewSp}
 \delta S_p &\simeq& 2\mu \int_\Gamma  \exd \tau \, \Bigl[
 \Bigl( h \, \dot t^2 + k \, g^{tt} \Bigr)
 \delta \varphi_{,\,t} + h \, \dot t \, \dot r \,
 \delta \varphi_{,\,r} \Bigr] \nn\\
 &\simeq& 2\mu \int_\Gamma  \exd \tau \, \Bigl[
 (h - k) \, \delta \varphi_{,\,t} + h \, \dot r \,
 \delta \varphi_{,\,r} + \O(v^2) \Bigr] \,,
\ea
where $\delta \varphi_{,\,r} := \partial_r \delta \varphi$ and $\delta \varphi_{,\,t} := \partial_t \delta \varphi$. Notice that the shift symmetry, $\varphi \to \varphi + \omega$, guarantees that $\delta \varphi$ always appears differentiated, and so in principle eq.~\pref{NewSp} gives a different response than would the expansion $m(\varphi) \simeq m_0 + m'_0 \, \delta \varphi + \cdots$ that dominates the response of other matter sources, like neutron stars.

Considerations like these show that computing the changes to black hole orbits due to an asymptotic scalar evolution can be complicated, even in the post-Newtonian approximation. However the same issues do not similarly complicate the rate that scalars are emitted by black holes in binary systems, provided we work to leading nontrivial order in $\mu$. This is because the black hole scalar charge is itself of order $\mu$, and so the leading contribution to dipole radiation comes from the radiation due to a charge of order $\mu$ moving along the zeroth-order Schwarzschild orbits. Although the $\O(\mu^2)$ corrections to these orbits are complicated to compute, they are subdominant contributions to the scalar dipole radiation.

\subsection{Dipole radiation from binaries}
\label{sec:bhbinfr}

Consider first the problem of how a single black hole radiates energy into the scalar field as a function of its trajectory, which we assume (temporarily) to be given. Although we assume in this section that the scalar is massless, it really suffices that it be light enough to have modes into which the system can radiate. Since periodic motion with angular frequency $\Omega$ typically radiates dominantly into modes with angular frequency $\Omega$, this requires $m c^2/\hbar \ll \Omega$ \cite{cloutier}.

\subsubsection*{Scalar radiation from a single particle moving on a given trajectory}

In order to find the power emitted into scalar radiation, the scalar field
far from the source is written as
\be
 \varphi(\bfx, t) = \mu t - \frac{G}{r 
 } \left[
 \psi + \frac{x^{i}}{r 
 } \,
 \psi_i' +
 \frac{x^{i}x^{j}}{2r^{2} 
 } \;
 \psi_{ij}'' + \ldots
 \right]
 + \mathcal{O}\left(\frac{1}{r^{2}}\right) \,,
\ee
where $\psi_{i_{1}\cdots i_{l}}(t)$ is the time-dependent STF (symmetric trace-free) multipole moment of
order $l$ \cite{multipole}.
Primes denote differentiation with respect to time and the right-hand side is to be evaluated at the retarded time, $t_r := t - r$, where $r := |\bfx|$ denotes the (large) distance from the source to the field point of interest.
The terms in the square brackets may be thought of as a series in $v$, the typical velocity
of the source. Terms that fall off as $1/r^{2}$ and faster do not contribute to the emitted power.
The contribution to the emitted power from the dipole moment $\psi_{i}$ is given by \cite{damourrev}
\be
 \label{dippwr1}
 F_{\varphi}^{\rm dip} = \frac{G}{3
 } \sum_{i}
 \left( {\psi_i''}
 \right)^{2} \,.
\ee

Explicitly, in terms of the non-relativistic source density, $\varrho(\bfx, t)$, defined by the field equation
\be
 \label{scsource}
 \Box_{f} \varphi = 4 \pi G \varrho 
 + \mathcal{O}(v^{4}
 ) \,,
\ee
where $v$ is the typical velocity of the source, and $\Box_{f}$ is the flat-space d'Alembertian operator, we have
\ba \label{dipmom}
 \psi(t_r) &=& \int \exd^{3}x
 \; \varrho(t_r, \bfx) +
 \mathcal{O}(v^{2}
 ) \nn\\
 \psi_{i}(t_r) &=& \int \exd^{3}x
 \; \varrho(t_r, \bfx) \, x_{i} +
 \mathcal{O}(v^{2}
 ) \,,
\ea
and so on.

For a point object with scalar charge $Q$ and trajectory $\bfz(t)$ we have $\varrho(t, \bfx) = Q \, \delta^{3}[\bfx - \bfz(t) ]$, and so $\psi = Q + \O(v^2)$ and $\psi_{i}(t) = Q \, z_i(t) + \O(v^2)$, {\em etc}. Equations (\ref{dippwr1}), (\ref{scsource}) and (\ref{dipmom}) show that the $Q$ that appears in standard calculations of the scalar dipole radiation rate is the same $Q$ that is identified in section \ref{sec:bh} using Jacobson's Miracle Hair Growth formula for asymptotically varying scalar fields. This allows the use of standard expressions for scalar radiation when computing the leading dipole scalar radiation rate for binary systems.

\subsubsection*{Scalar radiation from binary systems}

Next consider the radiation rate from a pair of compact objects having masses $M_{\ssA}, M_{\ssB}$, and scalar charges, $Q_\ssA$, $Q_\ssB$. An important role is played by the ratios $a_{\ssA} = Q_\ssA/M_\ssA, a_{\ssB} = Q_\ssB/M_\ssB$. As emphasized earlier, this ratio would be the asymptotic scalar-matter coupling, $a(\varphi_\infty)$, for ordinary stars in Brans-Dicke-like theories, but this need not be true for objects with relativistic structure, like neutron stars or black holes.

As usual, when all other things are equal, we expect scalar radiation to be well constrained by binary orbits because it can arise as dipole radiation. This is by contrast with gravitational radiation, which first appears at quadrupole order and so with a rate that is suppressed relative to dipole radiation by additional factors of $v$, with $v$ being a representative orbital speed (which for numerical purposes we take to be the average orbital speed).

To quantify this, let $F^{\rm mon}$, $F^{\rm dip}$ and $F^{\rm quad}$ respectively denote the power in monopole, dipole and quadrupole radiation, integrated over an orbital period. For objects whose orbits are close to Newtonian the total energy loss is the sum of tensor and scalar contributions, $F = F_g + F_\varphi$, with the leading gravitational contribution being quadrupole radiation, $F_g \simeq F_g^{\rm quad} + \cdots$. By contrast, for scalars monopole and dipole radiation is possible, $F_\varphi \simeq F_\varphi^{\rm mon} + F_\varphi^{\rm dip} + F_\varphi^{\rm quad} + \cdots$.

In order of magnitude $F_{\varphi}^{\rm mon} = \mathcal{O}[a_{\AcB}^{2} v^{5}]$, $F_{\varphi}^{\rm dip} = \mathcal{O}[a_{\AcB}^{2} v^{3}]$, $F^{\rm quad}_{g} = \mathcal{O}[v^{5}]$ and $F^{\rm quad}_{\varphi} = \mathcal{O}[a_{\AcB}^{2} v^{5}]$. (Notice that $F_{\varphi}^{\rm mon}$ is order $v^{5}$ rather than order $v$ because the scalar charges $Q_{\AcB}$ are time-independent.)
Constraints on scalar-tensor theories arise as upper bounds on the scalar radiation rate for systems where the gravitational rate is observed. To this end define the ratio
\be
 \xi := \frac{F_\varphi}{F_g} \simeq \frac{F^{\rm dip}_\varphi}{F^{\rm quad}_g} = \mathcal{O}\left[ \frac{a_{\AcB}^{2}}{v^2} \right] \,.
\ee

Explicit formulae for the orbitally averaged power emitted into scalar and gravity waves are calculated in chapter 6 of \cite{damourrev}, which finds (re-introducing the factors of $c$)
\begin{eqnarray}
\label{dippwr}
 F^{\rm dip}_\varphi &=& \frac{\nu^{2} (GM\Omega)^{8/3} (1+e^{2}/2)(a_{\ssA}-a_{\ssB})^{2}}
 {3c^{3}G(1-e^{2})^{5/2}} \Bigl(1 + \mathcal{O}[(v/c)^{2}]
 \Bigr) \,,
\\
\label{quadpwr}
 F^{\rm quad}_{g} &=& \frac{32 \nu^{2} (GM\Omega)^{10/3}(1+73e^{2}/24 + 37e^{4}/96)}
 {5c^{5}G(1-e^{2})^{7/2}} \Bigl(1  + \mathcal{O}[(v/c)^{2}] \Bigr) \,,
\end{eqnarray}
where $e$ is the orbit eccentricity, $\Omega = 2\pi/P$ where $P$ is the orbital period, $M=M_{\ssA}+M_{\ssB}$ is the total mass, and $\nu = M_{\ssA}M_{\ssB}/M^{2}$. Using these, the ratio $\xi$ becomes
\begin{eqnarray}
\label{qest1}
  \xi &=& \frac{5c^{2}(a_{\ssA}-a_{\ssB})^{2}(1-e^{2})(1+e^{2}/2)}
 {96(GM\Omega)^{2/3}(1+73e^{2}/24+37e^{4}/96)} \Bigl(1 + \mathcal{O}[(v/c)^{2}] \Bigr)
\nonumber
\\
\label{qest}
 &\simeq& \frac{5}{96}
 \, \frac{
 (a_{\ssA}-a_{\ssB})^{2}}{(v/c)^2} \; f(e)
 \Bigl( 1 + \mathcal{O}[(v/c)^{2}] \Bigr) \,.
\end{eqnarray}
Notice that $(GM\Omega/c^{3})^{2/3} = GM/\frak{a} c^2 = (v/c)^{2}$ in the Keplerian limit, where $\frak{a}$ is the semi-major axis of the orbit and $v$ is the average orbital speed.

For two black holes with an asymptotically static scalar field we have $a_\ssA = a_\ssB = 0$ because of the no-hair condition $Q_\ssA = Q_\ssB = 0$, as is well-known to hold for Kerr and Schwarzschild black holes in general relativity \cite{willzag}. This implies in particular the usual absence of dipole radiation, $\xi \simeq 0$, as expected.

When the asymptotic scalar varies slowly in time --- {\em i.e.} $\mu \ne 0$ --- we've seen that the above expressions can also be used to compute the leading contribution to the scalar dipole radiation, with the scalar charge given as in section \ref{sec:bh}: $a_\ssA = 2 \epsilon_\ssA$ and $a_\ssB = 2 \epsilon_\ssB$. Since both black holes see the same asymptotic scalar field far from the binary system, consistency requires
\begin{equation}
 \frac{\epsilon_{\ssA}}{2GM_{\ssA}} =
 \frac{\epsilon_{\ssB}}{2GM_{\ssB}} = \mu \,,
\end{equation}
and so because the two black holes share the same asymptotic time-dependent scalar field, their scalar charges are related to one another.

The quantity appearing in $\xi$ is $a_\ssA - a_\ssB = 2(\epsilon_\ssA - \epsilon_\ssB) = 4 G \mu (M_\ssA - M_\ssB)$. We see that inspiralling black holes with differing masses will radiate dipole radiation when situated within a slowly varying asymptotic scalar field, provided only that the scalar is light enough to receive the radiation. The amount of radiation is guaranteed to be small by virtue of the approximation $\epsilon_{\AcB} \ll 1$ that underlies our linearized analysis.

This calculation assumes $Q_\ssA$ and $M_\ssA$ are fixed quantities over any one orbit, and this is true in the present case if $\epsilon_\AcB \ll 1$ and $\mu$ doesn't change in time. Notice in particular that (at leading order) we need {\em not} assume $\mu \ll \Omega$ for the radiation-loss calculation to be valid.

\section{The Quasar OJ287}
\label{sec:oj287}

We next apply the formalism of section \ref{sec:bhbinfr} to OJ287. This is a quasar situated at redshift $z=0.306$ that undergoes periodic bursts. This system has been argued to consist of two gravitationally-bound black holes \cite{oj287, oj287GR, oj287nohair}, with the bursts occurring when the `small' black hole (black hole B, with mass $M_\ssB \simeq 10^8$ $M_\odot$) passes through the accretion disc of the larger one (black hole A, which is a monster with $M_\ssA \simeq 10^{10}$ $M_\odot$).

The orbital parameters of the system have been reconstructed from the observed bursting pattern, leading to the orbital parameters summarized in table \ref{orbpars}. In particular, the success of the orbital description requires the orbital decay due to the emission of gravitational radiation, with the observed radiated power being within $6\%$ of the prediction of General Relativity \cite{oj287GR}.

Taking this binary black hole interpretation as correct, we may use its success to infer an upper limit for the time-variation of any ambient light scalar fields that were present in the black hole neighborhood at an epoch $z = 0.306$. Notice that this does {\em not} require making any assumptions about the nature of the scalar-matter couplings. This bound applies to any scalar fields at this epoch, provided that they are light enough that it is possible to radiate into their scalar modes. This requires the mass of the scalar to be much lighter than the angular frequency, $\Omega$. Given that the orbital period is about 9 years, the corresponding scalar mass limit becomes $m \ll 10^{-23}$ eV/$c^2$ \cite{cloutier}.

Using the black hole orbital parameters in equation (\ref{qest}), we find $\xi \sim 0.94 (a_{\ssA}-a_{\ssB})^{2}$, and so imposing $\xi < 0.06$ gives the constraint $|a_{\ssA}-a_{\ssB}| < 0.25$ and so $|\epsilon_{\ssA}-\epsilon_{\ssB}| < 0.13$. We have $\epsilon_{\AcB} = 2GM_{\AcB} \, \mu/c^{3}$, where $2GM_{\ssA}/c^{3} \sim 1.8 \cdot 10^{5}\ {\rm s} \sim 2\ {\rm day}$ is the light-crossing time of black hole A, and $2GM_{\ssB}/c^{3} \sim 1.4 \cdot 10^{3}\ {\rm s} \sim 23\ {\rm min}$ is that of black hole B.

This finally implies $|\mu^{-1}| > 1.4 \cdot 10^{6}\ {\rm s} \sim 16\ {\rm days}$. This limit holds even though it is much smaller than the orbital period, because the instantaneous radiation rate relies only on the constancy of $\mu$ and the black hole masses and scalar charges. Our upper bound on $|\mu|$ implies $|\epsilon_{\ssA}| < 0.13$ and $|\epsilon_{\ssB}| < 1.0 \cdot 10^{-3}$. This is consistent with our linearization assumption $\epsilon_{\AcB} \ll 1$.

\begin{table}
\begin{tabular}{|l|l|l|}
\hline
Parameter & Meaning & Value \\
\hline
$\dot \omega$	& Precession rate						& $39.1(1)^{\circ}$ per orbit \\
$M_{\ssA}$		& Mass of black hole A 					& $1.84(1) \cdot 10^{10} M_{\odot}$ \\
$M_{\ssB}$		& Mass of black hole B 					& $1.40(3) \cdot 10^{8} M_{\odot}$ \\
$\chi_{\ssA}$	& Dimensionless Kerr parameter of black hole A 	 & $0.28(3)$ \\
$e$ 			& Orbital eccentricity					& $0.658(1)$ \\
$P$			& Orbital period						& $\sim 9\ {\rm years}  \sim 2.8 \cdot 10^{8} {\rm s}$ \\
$\frak{a}$		& Semi-major axis						& $\sim 1.7 \cdot 10^{15} {\rm m}$ \\
$v_{\rm orb}$	& Typical orbital velocity					 & $\sim 0.13c$ \\
\hline
\end{tabular}
\caption{Orbital parameters of OJ287.
These are the `intrinsic' values, as measured in the centre-of-mass of the binary. The values measured on Earth differ by a redshift factor. Data taken from \cite{oj287pars}. Uncertainties are given at the 3 sigma level.}
\label{orbpars}
\end{table}

Notice also that the spin of black hole $A$ is claimed to have been measured \cite{oj287spin}. The solution described in section \ref{sec:bh} can be generalized to rotating black holes \cite{jacobson}, and the analogue of equation
(\ref{efscmcoupl}) for Kerr black holes is
\begin{equation}
\label{scchkerr}
a_{\ssA} = 2GM_{\ssA} \mu (1 + \sqrt{1-\chi_{\ssA}^{2}}) \,,
\end{equation}
where $\chi_{\ssA} = cJ_{\ssA}/GM_{\ssA}^{2}$ is the dimensionless Kerr
parameter, and $J_{\ssA}$ is the angular momentum of the black hole.
It follows from equation (\ref{scchkerr}) that taking spin into account
does not significantly change our bounds.

\section{Summary and Outlook}
\label{sec:summary}

In this paper we use Jacobson's Miracle Hair-Growth Formula to show that inspiralling black holes can radiate dipole scalar radiation whenever there exists a scalar whose mass is small enough to be radiated (typically this requires the mass to be much smaller than the orbital period, $m \ll \Omega$). We then apply this radiation calculation to the quasar OJ287, which has been interpreted as a very massive double black hole binary. Using the inferred black hole masses and orbital properties, and the success of the GR prediction for the orbital decay rate, we conclude that any ambient scalar fields lighter than $m < 10^{-23}$ eV at this epoch (redshift $z = 0.306$) must have instantaneous time evolution over scales larger than $|\mu^{-1}| \gtrsim 16$ days.

Although this is not a spectacular bound compared with cosmological time evolution, there are two things about it that are quite remarkable. First, it is completely independent of how the light scalar couples to matter since this does not enter into the amount of acquired black hole hair. Second, it directly constrains the instantaneous scalar evolution rate, rather than an average change in scalar field values over cosmological time intervals.

This system is representative of the rich kinds of physics that can arise for gravity/scalar systems with time-dependent environments, and indicates the wealth of information that is likely to become available once gravitational radiation is detected from astrophysical sources.

Direct detection of gravitational waves in the future will provide a systematic means of measuring the properties of double black hole binaries, making it possible to map out the bounds on $|\mu^{-1}|$ at different redshifts. In preparation for these happy times it would be useful to predict in more detail the response of these systems to scalar-tensor theories, and in particular how their orbits respond to their scalar `environment'.

Alternatively, if the scalars are not precisely massless they will propagate with different speeds than those of the primary gravitational signal, allowing potential time delays in gravitational wave signals whose presence could signal the existence of a new scalar component to gravity.

\section*{Acknowledgements}

We thank Craig Heinke, Ted Jacoboson, Laura Parker, Ethan Vishniac and James Wadsley for helpful discussions. Our research was supported in part by funds from the Natural Sciences and Engineering Research Council (NSERC) of Canada. Research at the Perimeter Institute is supported in part by the Government of Canada through Industry Canada, and by the Province of Ontario through the Ministry of Research and Information (MRI).

\end{document}